# The validity of the Stokes-Einstein relation in ionic liquids

Gan Ren (任淦)[†], Tao Duan (段涛)

School of Science, Civil Aviation Flight University of China, Guanghan, 628307, China

**Abstract**

Ionic liquids (ILs) exhibit supercooled liquids behavior even above room temperature and the Stokes-Einstein (SE) relation is often considered to be invalid in ILs as that in supercooled liquids. However, the conclusion is usually drawn based on some variants of SE relation. In this work, we have systematically investigated the validity of the Stokes-Einstein relation in ILs by performing molecular dynamics simulations of coarse-grained $[EMI^+][NO_3^-]$ within the temperature range of 400-800 K. Both the original SE formulation, $D \sim T/\alpha$, and two commonly employed variants, $D \sim \tau^{-1}$ and $D \sim T/\eta$, were examined to assess their applicability and consistency. These three formulas yield distinctly different results, indicating that the two variants are not reliable substitutes for the original SE relation. The inconsistency between the $D \sim T/\eta$ and $D \sim T/\alpha$ suggests the fact that the *Cr* in Stokes's law varies with conditions. The breakdown of $D \sim T/\alpha$ arises from the ion correlations introduced by the strong electrostatic interactions. This proposition is further confirmed by simulations with $[VIO^{2+}][Tf_2N^-]_2$ and variably charged $[EMI^{q+}][NO_3^{q-}]$.

## 1. Introduction

Owing to their ionic constitution, ionic liquids (ILs) exhibit a suite of distinctive properties including low volatility, electrical conductivity, tunable solubility, and chemical stability that have garnered considerable attention for their promising applications,[1-4] particularly in the case of typical room-temperature ILs.[5, 6] The diverse and complex nature of ions in ILs enables the formation of rich microscopic structures and phases, including ionic liquid clusters,[7] nanoscale domain,[8] ionic liquid crystal[9] and even the mixed liquid-glass state.[10] Additionally, ILs display supercooled liquid

[†] Corresponding author. E-mail: rengan@alumni.itp.ac.cn

characteristics even above ambient temperature.[11, 12]

The supercooled liquid-like behavior of ILs has been identified in numerous experimental [13-19] and theoretical studies.[20-22] Dynamic heterogeneity, a salient feature of supercooled liquids, has been identified in ILs through non-Gaussian parameter and dynamic susceptibility analyses.[20, 21, 23] The structural relaxation of ILs deviates from the simple exponential decay characteristic of normal liquids, instead exhibiting stretched-exponential behavior.[11, 22, 24-26] Mobile particles in ILs also exhibit stronger spatial correlations and display collective diffusion,[8, 20] analogous to that observed in supercooled liquids.[27] A dynamical crossover is also observed in ILs,[26, 28] similar to those found in supercooled water.[29] An increasing characteristic time and length scale upon cooling have also been detected in ILs.[30, 31] Furthermore, the Stokes-Einstein (SE) relation has been found to break down in ILs.[11, 32]

The SE relation $D = k_{\mathrm{B}}T/C\eta r$ combines the Einstein relation $D = k_{\mathrm{B}}T/\alpha$ and Stokes' law $\alpha = C\eta r$, establishing a connection between the diffusion coefficient $D$ and viscosity $\eta$ for a particle moving through a viscous fluid, where $k_{\mathrm{B}}$ is the Boltzmann constant, $T$ the temperature, $\alpha$ the frictional coefficient, $r$ the effective hydrodynamic radius, and $C$ being a constant depending on the boundary condition.[33] Two variants of the SE relation, $D \sim T/\eta$ [34, 35] and $D \sim \tau^{-1}$,[11, 36] are commonly employed to examine its validity in both supercooled liquids and ILs, where "~" means proportional. The former is derived under the assumption that the $r$ remains constant. The latter arises from the self-intermediate scattering function when the displacement follows Gaussian statistics.[37]

Jung and coworkers[11, 31] reported a breakdown of $D \sim \tau^{-1}$ in ILs and which exhibits a fractional form as $D \sim \tau^{-\xi}$ with $\xi \neq 1.0$. They attributed this breakdown to dynamic heterogeneity induced decoupling of exchange and persistence times. Ludwig and coworkers[32] investigated the SE relation in ILs and IL/chloroform mixtures, finding that the $D \sim T/\eta$ breaks down for both pure ILs and IL/chloroform mixtures down to the miscibility gap. The fractional form of $D \sim T/\eta$ shown like $D/T \sim \eta^{-\xi}$ in ILs has been found to correlate with the breakdown of the Nernst-Einstein relation.[38, 39] Deviations from the SE relation have also been observed for neutral and charged small solutes in ILs,[40, 41] where the $r$ is typically assumed constant in $D \sim T/\eta$. However, the $r$ calculated by formula $r = k_B T / C\eta D$ actually increases with temperature and fails to reflect the true ion size.[42]

Although the breakdown of the SE relation in ILs has been demonstrated through its variants, no systematic investigation has been conducted to directly examine the SE relation in its original form alongside its variants. Since both $D \sim \tau^{-1}$ and $D \sim T/\eta$ rely on specific assumptions and conditions, and evidence suggests that these variants are only qualitatively consistent yet quantitatively divergent,[43] therefore the validity of the SE relation in ILs even above room temperature remains elusive and warrants reexamination. In this work, we first examine the validity of the SE relation and its two variants through molecular dynamics (MD) simulations of coarse-grained 1-ethyl-3-methyl-imidazolium nitrate ($[EMI^+][NO_3^-]$) over the temperature range of 400-800K. We then perform MD simulations of differently charged $[EMI^{q+}][NO_3^{q-}]$ and dimethy-viologen bis-(tetrafluorborate) ($[VIO^{2+}][Tf_2N^-]_2$) to further investigate the influence of polarization effects and variation in effective hydrodynamic radius on the SE relation. The remainder of this paper is organized as follows: Section 2 provides a brief description of the simulation details and analysis methods; the results and discussion are presented in Section 3; and Section 4 summarizes our conclusions.

## 2. Simulation details and analysis methods

### 2.1 Simulation details

In this work, we first employ a coarse-grained $[EMI^+][NO_3^-]$ model[44] to investigate the SE relation and its variants. The simulated system comprises 1024 ion pairs in a cubic box with a side length of 6.31 nm. MD simulations were performed at ten temperatures spanning 400-800 K. To further examine the possible influence of polarization on the validity of the SE relation, we simulated the coarse-grained $[EMI^{q+}][NO_3^{q-}]$ system with only varying charges, where $q$ ranges from 0.5 to 1.2 in increments of 0.1. For the variable-charge model $[EMI^{q+}][NO_3^{q-}]$, the Lennard-Jones parameters were deliberately kept identical to the reference model $[EMI^+][NO_3^-]$. By varying only the partial charges while fixing the Lennard-Jones parameters, we can unambiguously attribute observed differences in transport properties to changes in Coulomb interactions, rather than to concomitant alterations in short-range interactions. Additionally, an all-atom MD simulation of $[VIO^{2+}][Tf_2N^-]_2$[45] was conducted from 500 to 800 K, containing 256 cations and 512 anions in a cubic box of 6.24 nm. All simulations were performed using the GROMACS package.[46, 47] The system temperature was maintained constant via the Nosé-Hoover thermostat.[48, 49] Periodic boundary conditions were applied in all three dimensions. Long-range electrostatic interactions were calculated using the particle mesh Ewald method[50] with a cutoff of 1.4 nm, and van der Waals interactions were treated with the same cutoff.

**2.2 Analysis methods**

To investigate the SE relation and its variants, the diffusion coefficient $D$ was calculated from its asymptotic relationship with the mean square displacement

$$D = \lim_{t\to\infty} \left\langle \sum_{i}^{N} \left| \vec{r}_i(t) - \vec{r}_i(0) \right|^2 \right\rangle \Big/ 6Nt \quad (1)$$

where $N$ is the number of ion, $\vec{r}_i(t)$ is the position of $i$th ion at time $t$, < > means an ensemble average. The structural relaxation time $\tau$[51] was determined from the self-intermediate scattering function

$$F_s(k,\tau) = e^{-1} \quad (2)$$

where $F_s(k,t)$ is defined by $F_s(k,t) = \sum_{j=1}^{N} \left\langle e^{\mathrm{i}k\cdot\left[\vec{\mathbf{r}}_j(t) - \vec{\mathbf{r}}_j(0)\right]} \right\rangle \Big/ N$; the $k$ is 10.5 $nm^{-1}$ for $[EMI^+]$ and 11.0 $nm^{-1}$ for $[NO_3^-]$, which corresponds to the position of the first peak in the static structure factor.

We adopt the method proposed by Hess to determine the shear viscosity $\eta$, owing to its reliability and rapid convergence.[52] An external force $a_x = A\cdot\cos(pz)$ is applied in the $X$ direction; upon reaching a non-equilibrium steady state, the shear viscosity can be obtained from

$$\eta = A\rho \big/ Vp^2 \quad (3)$$

where $A$ is the maximum of $a_x$, $p = 2\pi/l$ with the simulation box size $l$, $V$ the maximum of velocity in $X$ direction. Because the density $\rho$ and charge $p$ remain constant for each ionic liquid across different temperatures, the shear viscosity is evaluated as $\eta \sim A/V$ in our work. To ensure good precision, five acceleration values $A$ (ranging from 0.005 to 0.025 $nm/ps^2$) were employed to calculate the viscosity within the linear response regime.

The frictional coefficient $\alpha$ was determined by applying a static external electric field $E$ to the system. Upon reaching a non-equilibrium steady state, the frictional force ( $f_r = \alpha v$ ) on an ion balances the applied force ($qE$), the $\alpha$ is obtained from

$$\alpha = \frac{qE}{v} \quad (4)$$

where $v$ is calculated via $v = \lim_{t \to \infty} < r(t) > / t$ [44]. The charge $q$ remains constant for each ion, and $\alpha$ is evaluated as $\alpha \sim E/v$ . To ensure high precision, seven electric field strengths $E$ (ranging from 0.01 to 0.07 V/nm) were employed to determine the $\alpha$ within the linear response regime. As an illustration, the linear response regimes for various charged systems are presented in Fig. 1.

Dynamic heterogeneity is characterized by the non-Gaussian parameter,[51] defined as

$$\alpha_2(t) = 3\langle r^4(t)\rangle / 5\langle r^2(t)\rangle^2 - 1 \quad (5)$$

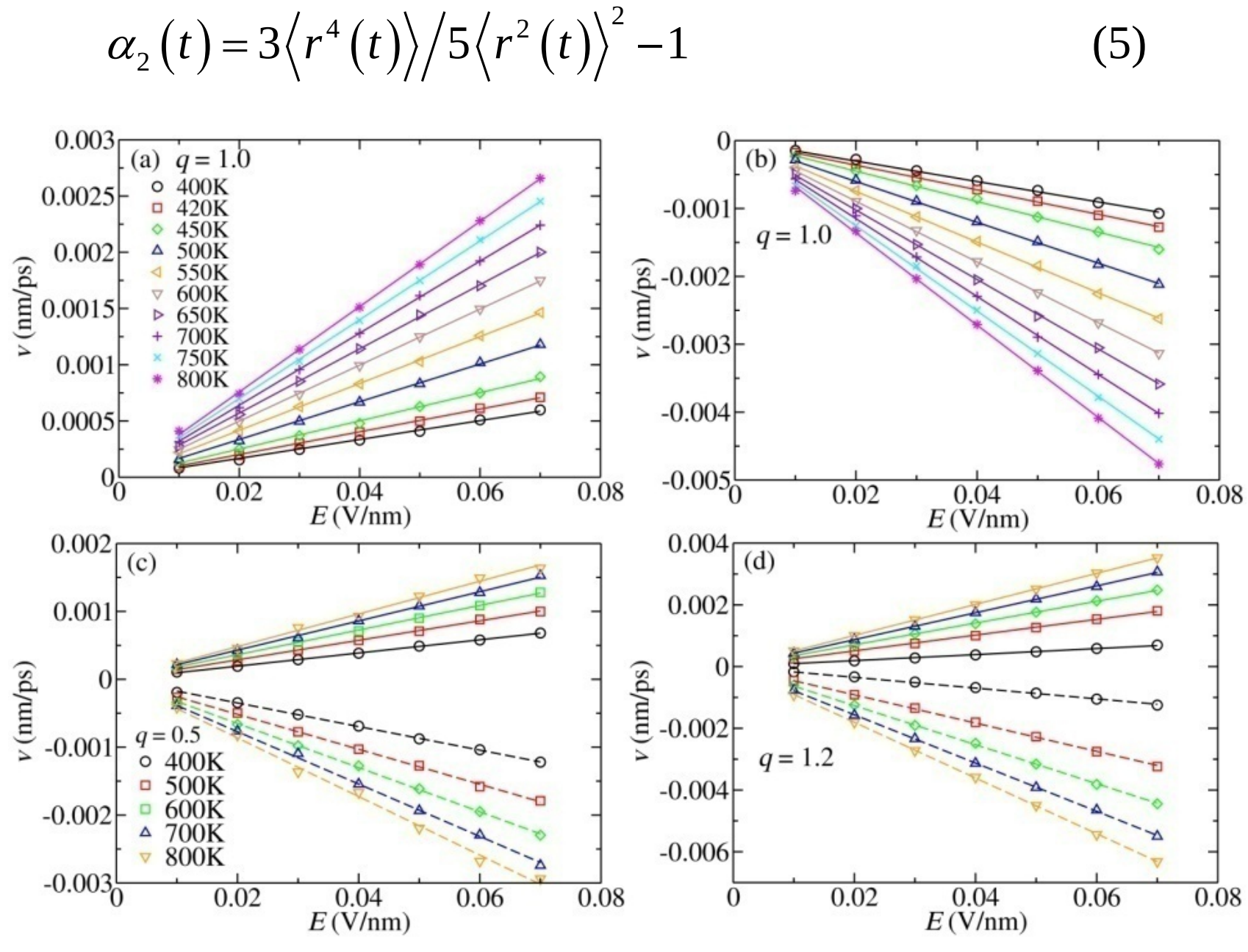

**Fig.1.** The linear response regimes for [$EMI^+$] (a), [$NO_3^-$] (b), [$EMI^{0.5+}$][$NO_3^{0.5-}$] (c), [$EMI^{1.2+}$][$NO_3^{1.2-}$] (d).

## 3. Results and discussion

To investigate the SE relation and its two variants in ILs, the calculated $D$, $\tau$, $\eta$ and $\alpha$ for [$EMI^+$] and [$NO_3^-$] as a function of $T$ are plotted in Fig. 2. As the variants typically exhibit fractional form, we employ the following two formulas $D \sim \tau^{-\xi_1}$ and $D \sim (T/\eta)^{\xi_2}$ to examine the two variants, and take the formula $D \sim (T/\alpha)^{\xi_3}$ to examine the original SE relation. A variant or the original SE relation is valid if the corresponding exponent $\xi_i = 1.0$ ($i$=1, 2 or 3), and invalid otherwise. Using the data presented in Fig.2, the original SE relation and its two variants are shown in Fig. 3.

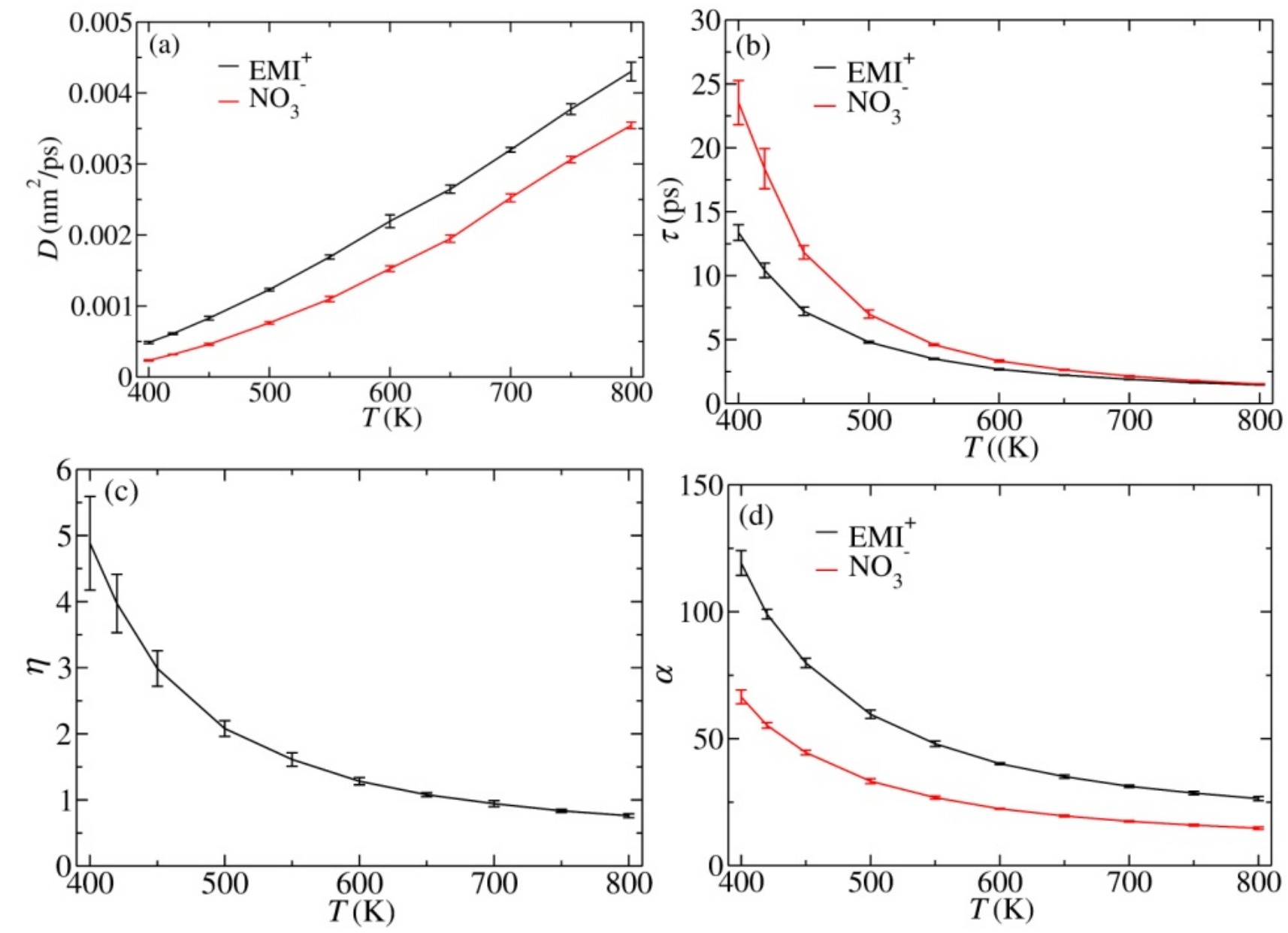

**Fig.2.** The $D$, $\tau$, $\eta$ and $\alpha$ for [$EMI^+$] and [$NO_3^-$] as a function of $T$: (a) $D$ vs $T$; (b) $\tau$ vs $T$; (c) $\eta$ vs $T$; (d) $\alpha$ vs $T$.

The fitted $\xi_1$ for both [$EMI^+$] and [$NO_3^-$] are approximately equal to 1.0, indicating the validity of the variant $D \sim \tau^{-1}$ within the simulated temperature range of 400-800K. The $D \sim \tau^{-1}$ holds exactly when the particle displacements follow Gaussian distribution, for which the $F_s(k,t)$ can be expressed as $F_s(k,t) = e^{-k^2 Dt}$. This is consistent with the non-Gaussian parameter $\alpha_2(t)$ shown in Fig. 4. Although the system exhibits increasing deviation from Gaussian behavior as temperature decreases, the maximum of $\alpha_2(t)$ remains small for both [$EMI^+$] and [$NO_3^-$] even at 400 K.

Our result differs from that reported in a coarse-grained MD simulation of [$EMI^+$][$PF_6^-$][11] over the temperature range of 300–800 K, where $D \sim \tau^{-1}$ fails for both [$EMI^+$] and [$PF_6^-$], with $\xi_1$ of 0.87 and 0.92, respectively. The [$PF_6^-$] anion is substantially heavier than [$NO_3^-$]; correspondingly, [$EMI^+$][$PF_6^-$] exhibits slower dynamics than [$EMI^+$][$NO_3^-$] at the same temperature, as reflected in the longer $\tau$. Moreover, previous studies[31] have shown that the breakdown temperature and exponents are closely correlated with the ionic charge in ILs: the $D \sim \tau^{-1}$ breaks down much earlier in charged systems than in their neutral counterparts, with the charged systems exhibiting a smaller $\xi_1$ compared to the neutral ones. Atomistic MD simulations with [$VIO^{2+}$][$Tf_2N^-$] report a similar trend.[53] Although [$VIO^{2+}$] carries twice the charge of [$EMI^+$], its larger ionic size partially compensates for this effect. Specifically, the SE relation is nearly perfectly satisfied for [$Tf_2N^-$] ($\xi_1$=1.01), whereas a slight breakdown is observed for [$VIO^{2+}$] ($\xi_1$=0.93) in the temperature range of 400–800 K. The latter is attributed to the more heterogeneous dynamics of [$VIO^{2+}$] arising from its rigid and extended molecular structure. Taken together, the validity of the SE relation observed in our work can be attributed to the relatively weak dynamic heterogeneity in [$EMI^+$][$NO_3^-$] within the simulated temperature range as shown in Fig. 4. We anticipate that a more pronounced breakdown may emerge at sufficiently lower temperatures, where dynamic heterogeneity is expected to intensify.

The $D \sim T/\eta$ depicted in Fig. 3b yields $\xi_2 = 0.852$ for [$EMI^+$] and 1.061 for [$NO_3^-$], suggesting the validity of $D \sim T/\eta$ for the [$NO_3^-$] but breakdown for the [$EMI^+$]. Conversely, $D \sim T/\alpha$ plotted in Fig. 3c displays the reverse behavior: satisfactory agreement for [$EMI^+$] ($\xi_3 = 0.991$) but clear violation for [$NO_3^-$] ($\xi_3 = 1.235$). Taken together, the three relations yield distinct fitted exponents, indicating that neither variant serves as a reliable substitute for the original SE relation. Moreover, the differences between $D \sim T/\alpha$ and $D \sim T/\eta$ for [$EMI^+$] and [$NO_3^-$] suggest the *Cr* in Stokes's law is varied with temperature. The variation of *Cr* obeys $Cr \sim D^{(\xi_3 - \xi_2)/\xi_3\xi_2}$ by a combination of $D \sim (T/\eta)^{\xi_2}$ and $D \sim (T/\alpha)^{\xi_3}$, indicating that the *Cr* for [$EMI^+$] and [$NO_3^-$] decrease with decreasing temperature. This trend is consistent with the collective motion previously observed in ILs.[8, 20] However, both $D \sim T/\eta$ and $D \sim T/\alpha$ exhibit a mixed valid–invalid pattern, with the two formulas yielding seemingly contradictory results. What is the origin of this inconsistency?

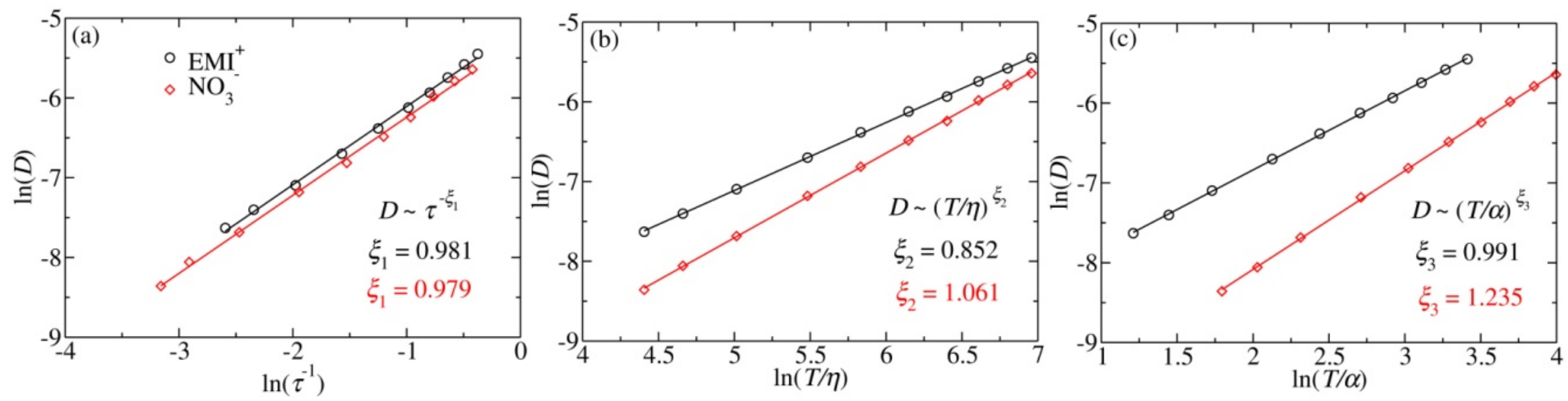

**Fig. 3.** Verification of the validities of the SE relation and its variants for [$EMI^+$][$NO_3^-$]: (a) $D \sim \tau^{-1}$; (b) $D \sim T/\eta$; (c) $D \sim T/\alpha$. The calculated data are represented by symbols and fitted by $D \sim \tau^{-\xi_1}$, $D \sim (T/\eta)^{\xi_2}$ and $D \sim (T/\alpha)^{\xi_3}$, respectively. The solid line and fitted exponent $\xi_i$ for [$EMI^+$] are in black and red for [$NO_3^-$].

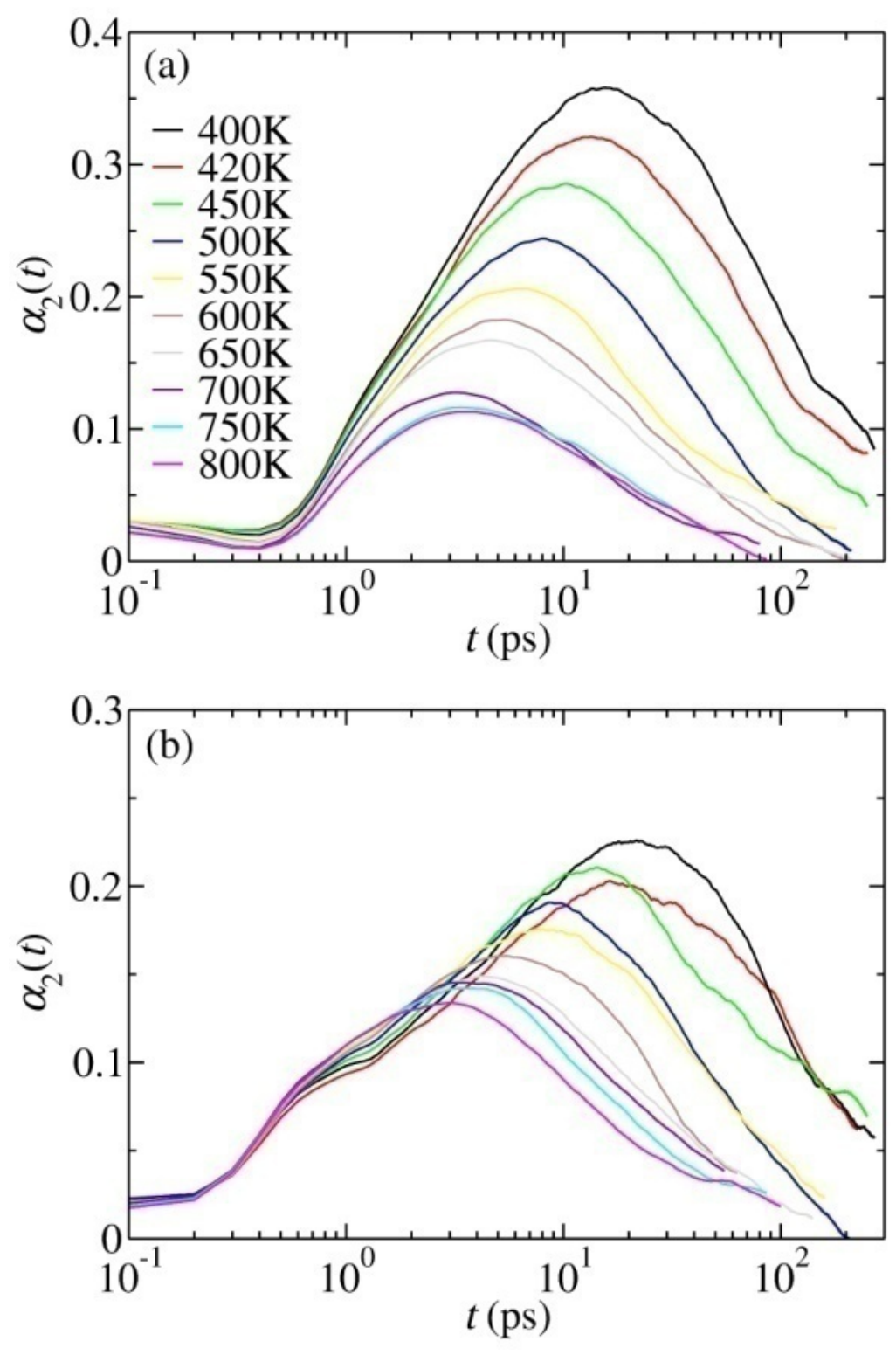

**Fig. 4.** The non-Gaussian parameter $\alpha_2(t)$ under different $T$ for $EMI^+$ (a) and $NO_3^-$ (b).

The electrostatic interactions in ionic liquids (ILs) are very strong. Previous studies have shown that non-exponential relaxation, dynamic heterogeneity, and the validity of the SE relation are closely correlated with the charges and charge distribution.[11, 21, 31] Furthermore, the SE relation holds in pure methanol but breaks down in mixtures with ILs.[54] Moreover, the Nernst-Einstein (NE) relation is found to break down in ILs; this breakdown arises from ion correlations introduced by strong electrostatic interactions, which introduce strong correlations in ionic motion and thus violate the prerequisite for the validity of the NE relation.[38] Since the NE relation is derived from the SE relation, the breakdown of the SE relation must likewise originate from strong correlations introduced by electrostatic interactions. As electrostatic interaction strengthens with increasing ion charge and diminished by more delocalized charge distributions, the $D \sim T/\alpha$ should become valid for a less charged [$NO_3^{q-}$] and for ILs with more delocalized charge distributions. To validate this hypothesis and obtain a quantitative understanding of electrostatic interaction on $D \sim T/\alpha$, we performed additional MD simulations on [$EMI^{q+}$][$NO_3^{q-}$] ($q$ = 0.5-1.2) and [$VIO^{2+}$][$Tf_2N^-$]$_2$. The temperature-dependent $D$, $\eta$ and $\alpha$ for [$EMI^{q+}$] and [$NO_3^{q-}$] are presented in Fig. 5; representative $D \sim T/\eta$ and $D \sim T/\alpha$ results for $q$ = 0.5, 0.8, and 1.1 appear in Fig. 6. Complete fitted exponents are provided in Table 1.

Similar to [$EMI^+$][$NO_3^-$], the $D \sim T/\eta$ remains almost valid for [$NO_3^{q-}$] across different charge values but fails for [$EMI^{q+}$]. However, the fitted exponent $\xi_2$ almost increases monotonically with increasing $q$, implying that the near validity of $D \sim T/\eta$ for [$NO_3^-$] shown in Fig. 3 is merely coincidental. Meanwhile, the fitted exponent $\xi_2$ is almost increased with decreasing $q$ and approaches 1.0, suggesting that $D \sim T/\eta$ may become valid for [$EMI^{q+}$] at sufficiently low charge, which also imply the coincidental near validity $D \sim T/\eta$ for [$NO_3^-$]. The results indicate that variations in the $Cr$ for [$EMI^{q+}$] and [$NO_3^{q-}$] are strongly correlated with electrostatic interaction. Combining the formula $Cr \sim D^{(\xi_3-\xi_2)/\xi_3\xi_2}$ and exponents in Table 1, we observe that $Cr$ exhibits more pronounced temperature dependence at higher charge, especially for [$NO_3^{q-}$]. The $D \sim T/\alpha$ is also valid for [$EMI^{q+}$] across different charge values but fails for [$NO_3^{q-}$]. However, the fitted exponent $\xi_3$ decreases and approaches 1.0 as $q$ decreases, indicating that $D \sim T/\alpha$ may become valid for [$NO_3^{q-}$] at sufficiently low charge. Moreover, the validity of $D \sim T/\alpha$ for [$EMI^{q+}$] arises from the delocalized charge distribution and the resulting weak electrostatic interactions. These results support our proposition that correlation effects introduced by electrostatic interaction weaken with decreasing charge, such that the breakdown of $D \sim T/\alpha$ vanishes in the low-charge limit.

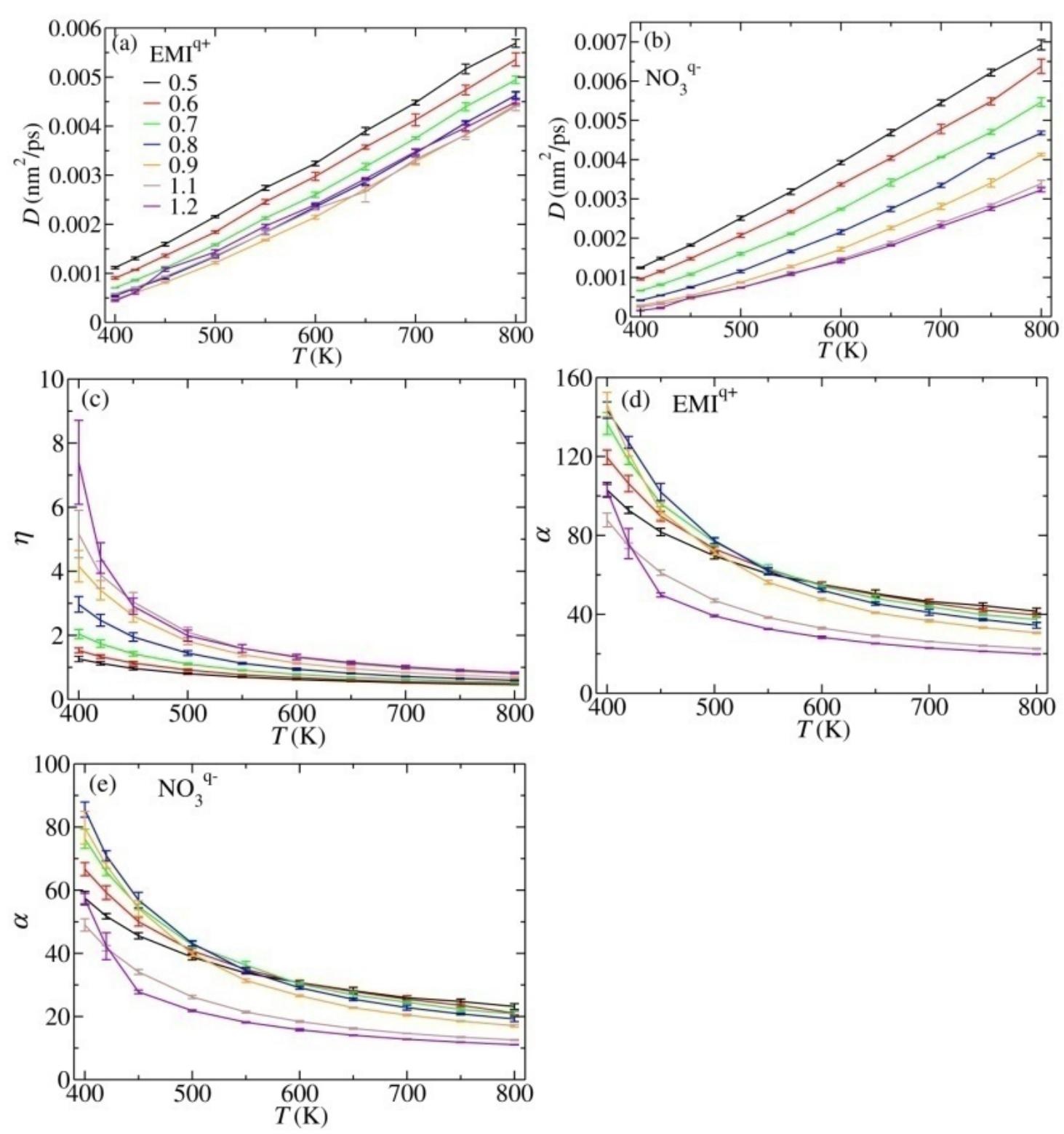

**Fig. 5.** The $D$, $\eta$ and $\alpha$ for different charged [$EMI^{q+}$] and [$NO_3^{q-}$] as a function of $T$: (a) $D$ for [$EMI^{q+}$] vs $T$; (b) $D$ for [$NO_3^{q-}$] vs $T$; (c) $\eta$ vs $T$; (d) $\alpha$ for [$EMI^{q+}$] vs $T$; (e) $\alpha$ for [$NO_3^{q-}$] vs $T$.

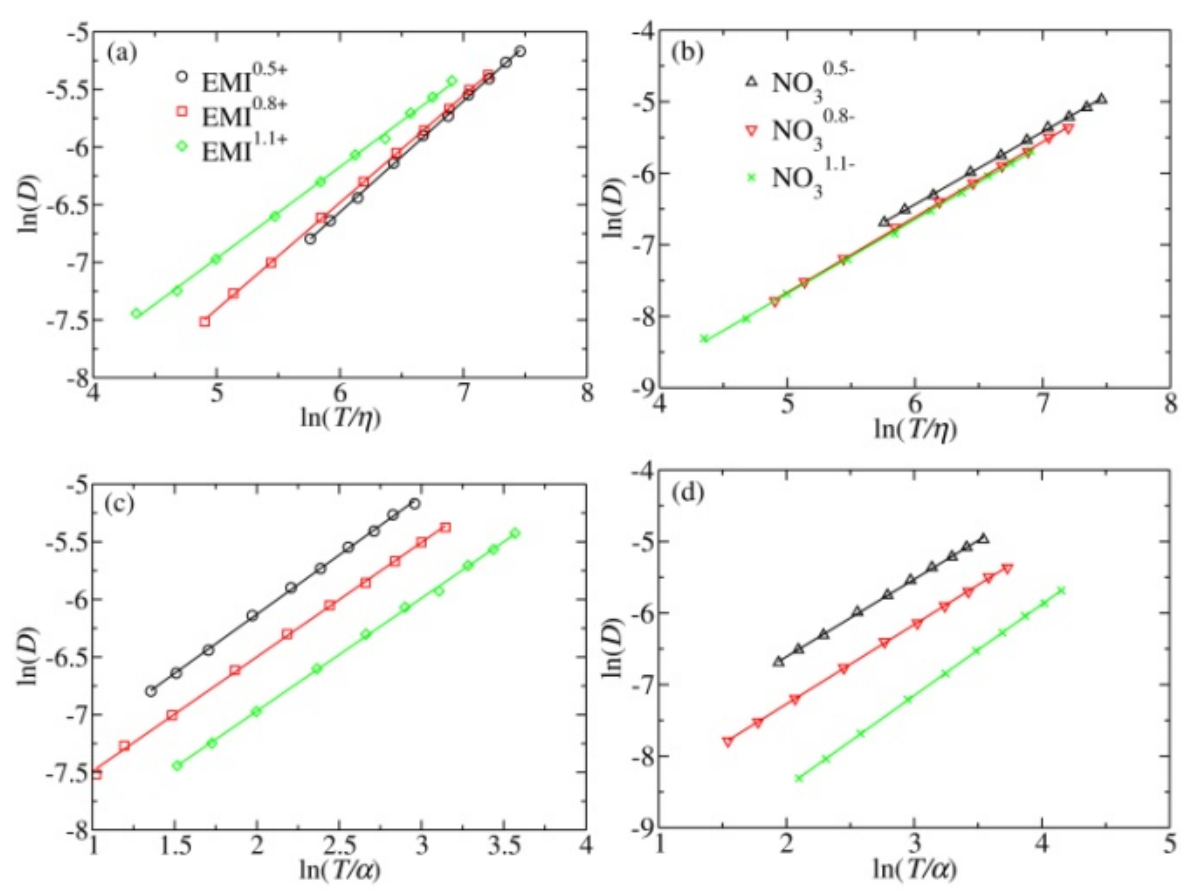

**Fig. 6.** Verification of the validities of $D \sim T/\eta$, $D \sim T/\alpha$ for different changed [EMI$^{q+}$] and [$NO_3^{q-}$], respectively: (a) $D \sim T/\eta$ for [EMI$^{q+}$]; (b) $D \sim T/\eta$ for [$NO_3^{q-}$]; (c) $D \sim T/\alpha$ for [EMI$^{q+}$]; (d) $D \sim T/\alpha$ for [$NO_3^{q-}$]. The calculated data are represented by symbols and fitted by $D \sim (T/\eta)^{\xi_2}$ and $D \sim (T/\alpha)^{\xi_3}$, respectively. The fitted exponents $\xi_i$ are listed in Table 1.

**Table 1.** The fitted $\xi_i$ in $D \sim (T/\eta)^{\xi_2}$ and $D \sim (T/\alpha)^{\xi_3}$ for different changed [EMI$^{q+}$] and [$NO_3^{q-}$].

| $q$ | 0.5 | 0.6 | 0.7 | 0.8 | 0.9 | 1.0 | 1.1 | 1.2 |
|---|---|---|---|---|---|---|---|---|
| $\xi_2^{[EMI^{q+}]}$ | 0.963 | 0.968 | 0.957 | 0.928 | 0.886 | 0.852 | 0.795 | 0.822 |
| $\xi_2^{[NO_3^{q-}]}$ | 1.015 | 1.020 | 1.035 | 1.052 | 1.058 | 1.061 | 1.036 | 1.083 |
| $\xi_3^{[EMI^{q+}]}$ | 1.027 | 0.994 | 0.980 | 0.993 | 0.990 | 0.991 | 0.979 | 0.997 |
| $\xi_3^{[NO_3^{q-}]}$ | 1.082 | 1.037 | 1.058 | 1.110 | 1.182 | 1.235 | 1.276 | 1.314 |

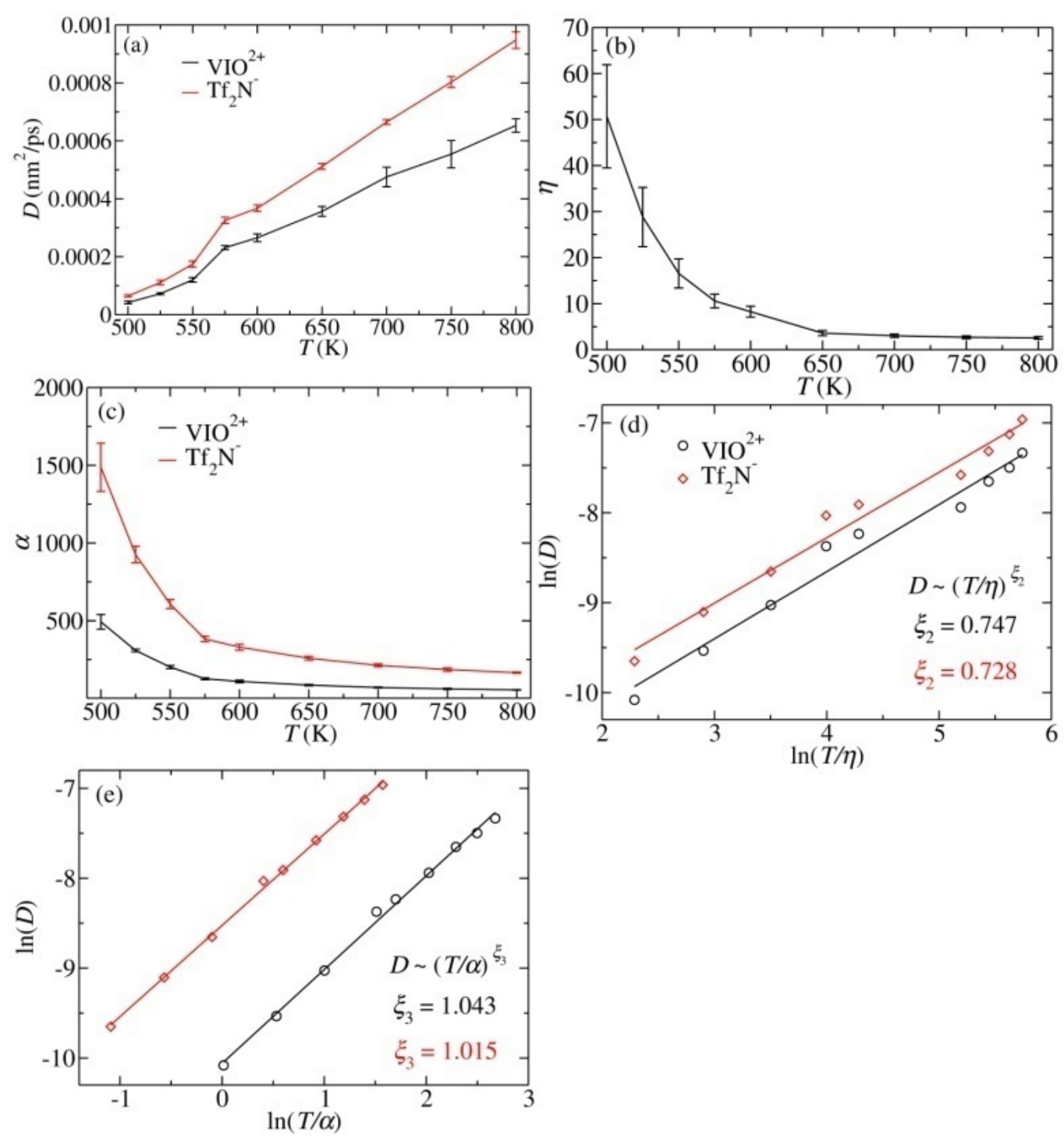

**Fig. 7.** The $D$, $\eta$ and $\alpha$ for [$VIO^{2+}$] and [$Tf_2N^-$] as a function of $T$: (a) $D$ vs $T$; (b) $\eta$ vs $T$; (c) $\alpha$ vs $T$. And verification of the validities of $D \sim T/\eta$, $D \sim T/\alpha$ for [$VIO^{2+}$] and [$Tf_2N^-$]: (c) $D \sim T/\eta$; (d) $D \sim T/\alpha$. The calculated data are represented by symbols and fitted by $D \sim (T/\eta)^{\xi_2}$ and $D \sim (T/\alpha)^{\xi_3}$, respectively. The fitted exponent $\xi_i$ is in the same color as the corresponding fitting line.

The calculated $D$, $\eta$ and $\alpha$ for [$VIO^{2+}$] and [$Tf_2N^-$], together with the SE relation analysis, are presented in Fig. 7. The fitted $\xi_2 = 0.747$ for [$VIO^{2+}$] and 0.728 for [$Tf_2N^-$] indicate that $D \sim T/\eta$ is invalid for this system. By contrast, the $\xi_3 = 1.043$ for [$VIO^{2+}$] and 1.015 for [$Tf_2N^-$] demonstrate that the original SE relation $D \sim T/\alpha$ remains established. The differences between $\xi_3$ and $\xi_2$ further suggests that the *Cr* for both [$VIO^{2+}$] and [$Tf_2N^-$] decreases with increasing temperature. Consequently, caution is warranted when employing the $D \sim T/\eta$ to assess SE relation validity. These findings corroborate our hypothesis: the original SE relation $D \sim T/\alpha$ is explicitly satisfied in the large-sized, charge-diffuse [$VIO^{2+}$][$Tf_2N^-$]$_2$ system with a weak electrostatic interaction.

## 4. Conclusion

In this work, we performed coarse-grained MD simulations of $[EMI^{q+}][NO_3^{q-}]$ from 400 to 800 K and atomistic MD simulations of $[VIO^{2+}][Tf_2N^-]_2$ from 500 to 800 K to investigate the validity of the SE relation and its variants in ILs above room temperature, as well as the consistency between the original SE relation and its variants. Our results demonstrate that $D \sim \tau^{-1}$, $D \sim T/\eta$ and $D \sim T/\alpha$ yield distinct outcomes, indicating that neither $D \sim \tau^{-1}$ nor $D \sim T/\eta$ serves as a reliable substitute for the original SE relation. The $D \sim \tau^{-1}$ remains almost valid for $[EMI^+][NO_3^-]$ over 400-800 K due to small deviations from Gaussian statistics in ion displacement. However, a mixed valid-invalid pattern and apparent inconsistencies emerge for $D \sim T/\eta$ and $D \sim T/\alpha$ in $[EMI^+][NO_3^-]$. Specifically, the observed breakdown of $D \sim T/\alpha$ for $[NO_3^-]$ arises from strong correlation effects introduced by electrostatic interaction. The *Cr* varies with temperature for both $[EMI^+]$ and $[NO_3^-]$, which critically influences SE relation testing via $D \sim T/\eta$. Our extended simulations corroborate these findings: the breakdown of $D \sim T/\alpha$ in $[NO_3^{q-}]$ vanishes for low-charge $[EMI^{q+}][NO_3^{q-}]$ and for the charge-delocalized $[VIO^{2+}][Tf_2N^-]_2$ system. Notably, *Cr* varies with temperature for both $[VIO^{2+}]$ and $[Tf_2N^-]$, as well as variation with charge for $[EMI^{q+}]$ and $[NO_3^{q-}]$, necessitating careful consideration of its conditions dependence when testing the SE relation. Our results confirm the conclusion that the validity of the original SE relation in ILs is closely correlated with the adopted form of the SE relation, underscoring the importance of electrostatic interactions in the validity of original SE relation and variations of *Cr* in Stokes's formula. These factors must be rigorously accounted for to avoid erroneous conclusions.

Despite being demonstrated for ILs, analogous phenomena may also be anticipated in aqueous ionic solutions and supercooled liquids. In electrolyte solutions, ions exist as solvated species surrounded by structured solvent shells and ion cages, exhibiting strong electrostatic interactions with the surrounding environment analogous to those in ILs. In pure molecular liquids, although intermolecular correlations are weak at high temperatures, they become strengthened once intermolecular interactions exceed thermal kinetic energy. Under such conditions, neither ions nor molecules move freely, potentially leading to a breakdown of the SE relation. Notably, ionic solutions may exhibit similar phenomena even above room temperature due to their strong electrostatic interactions.

**Acknowledgements**

This work was supported by the National Natural Science Foundation of China (No. 12104502) and the Fundamental Research Funds for the Central Universities (No. 26CAFUC03049, 25CAFUC09019).

## References

[1] Rogers R D and Seddon K R 2003 *Science* 302 792

[2] Plechkova N V and Seddon K R 2008 *Chem. Soc. Rev.* 37 123

[3] Mohammad A and Inamuddin D 2012 *Green Solvents II: Properties and Applications of Ionic Liquids* Springer)

[4] Armand M, Endres F, MacFarlane D R, Ohno H and Scrosati B 2009 *Nat. Mater.* 8 621

[5] Welton T 1999 *Chem. Rev.* 99 2071

[6] Wilkes J S, Levisky J A, Wilson R A and Hussey C L 1982 *Inorg. Chem.* 21 1263

[7] Chen S, Zhang S, Liu X, Wang J, Wang J, Dong K, Sun J and Xu B 2014 *Phys Chem Chem Phys* 16 5893

[8] Wang Y and Voth G A 2006 *J. Phys. Chem. B* 110 18601

[9] Ji Y, Shi R, Wang Y and Saielli G 2013 *J. Phys. Chem. B* 117 1104

[10] Ramírez-González P E, Sanchéz-Díaz L E, Medina-Noyola M and Wang Y 2016 *J. Chem. Phys.* 145 191101

[11] Jeong D, Choi M Y, Kim H J and Jung Y 2010 *Phys. Chem. Chem. Phys.* 12 2001

[12] Habasaki J, Leon C and Ngai K 2017 *Dynamics of glassy, crystalline and liquid ionic conductors* (Berlin: Springer)

[13] Hayamizu K, Tsuzuki S, Seki S and Umebayashi Y 2011 *J. Chem. Phys.* 135 084505

[14] Alam T M, Dreyer D R, Bielawski C W and Ruoff R S 2013 *J. Phys. Chem. B* 117 1967

[15] Cang H, Li J and Fayer M D 2003 *J. Chem. Phys.* 119 13017

[16] Lang B, Angulo G and Vauthey E 2006 *J. Phys. Chem. A* 110 7028

[17] Castner E W, Wishart J F and Shirota H 2007 *Acc. Chem. Res.* 40 1217

[18] Funston A M, Fadeeva T A, Wishart J F and Castner E W 2007 *J. Phys. Chem. B* 111 4963

[19] Khudozhitkov A E, Stange P, Bonsa A-M, Overbeck V, Appelhagen A, Stepanov A G, Kolokolov D I, Paschek D and Ludwig R 2018 *Chem. Commun.* 54 3098

[20] Del Pópolo M G and Voth G A 2004 *J. Phys. Chem. B* 108 1744

[21] Kim D, Jeong D and Jung Y 2014 *Phys. Chem. Chem. Phys.* 16 19712

[22] Arzhantsev S, Jin H, Baker G A and Maroncelli M 2007 *J. Phys. Chem. B* 111 4978

[23] Sang G and Ren G 2018 *J. Mol. Model.* 24 240

[24] Samanta A 2006 *J. Phys. Chem. B* 110 13704

[25] Shim Y, Jeong D, Manjari S, Choi M Y and Kim H J 2007 *Acc. Chem. Res.* 40 1130

[26] Habasaki J and Ngai K L 2008 *Anal. Sci.* 24 1321

[27] Donati C, Douglas J F, Kob W, Plimpton S J, Poole P H and Glotzer S C 1998 *Phys. Rev. Lett.* 80 2338
[28] Ren G 2021 *Chin. Phys. B* 30 16105
[29] Shi R, Russo J and Tanaka H 2018 *Proc. Natl. Acad. Sci. U. S. A.* 115 9444
[30] Kim S, Park S-W and Jung Y 2016 *Phys. Chem. Chem. Phys.* 18 6486
[31] Park S-W, Kim S and Jung Y 2015 *Phys. Chem. Chem. Phys.* 17 29281
[32] Köddermann T, Ludwig R and Paschek D 2008 *ChemPhysChem* 9 1851
[33] Landau L D and Lifshitz E M 1987 *Fluid Mechanics* (Oxford: Pergamon)
[34] Kumar P, Buldyrev S V, Becker S R, Poole P H, Starr F W and Stanley H E 2007 *Proc. Natl. Acad. Sci. U. S. A.* 104 9575
[35] Mazza M G, Giovambattista N, Stanley H E and Starr F W 2007 *Phys. Rev. E* 76 031203
[36] Ikeda A and Miyazaki K 2011 *Phys. Rev. Lett.* 106 015701
[37] Binder K and Kob W 2011 *Glassy materials and disordered solids: An introduction to their statistical mechanics* World Scientific)
[38] Harris K R 2010 *J. Phys. Chem. B* 114 9572
[39] Harris K R and Kanakubo M 2016 *Journal of Chemical & Engineering Data* 61 2399
[40] Araque J C, Yadav S K, Shadeck M, Maroncelli M and Margulis C J 2015 *J. Phys. Chem. B* 119 7015
[41] Kaintz A, Baker G, Benesi A and Maroncelli M 2013 *J. Phys. Chem. B* 117 11697
[42] Taylor A W, Licence P and Abbott A P 2011 *Phys. Chem. Chem. Phys.* 13 10147
[43] Shi Z, Debenedetti P G and Stillinger F H 2013 *J. Chem. Phys.* 138 12A526
[44] Shi R and Wang Y 2013 *J. Phys. Chem. B* 117 5102
[45] Ramírez-González P E, Ren G, Saielli G and Wang Y 2016 *J. Phys. Chem. B* 120 5678
[46] Berendsen H J, van der Spoel D and van Drunen R 1995 *Comput. Phys. Commun.* 91 43
[47] Van Der Spoel D, Lindahl E, Hess B, Groenhof G, Mark A E and Berendsen H J 2005 *J. Comput. Chem.* 26 1701
[48] Nosé S 1984 *J. Chem. Phys.* 81 511
[49] Hoover W G 1985 *Phys. Rev. A* 31 1695
[50] Darden T, York D and Pedersen L 1993 *J. Chem. Phys.* 98 10089
[51] Kob W, Donati C, Plimpton S J, Poole P H and Glotzer S C 1997 *Phys. Rev. Lett.* 79 2827
[52] Hess B 2002 *J. Chem. Phys.* 116 209
[53] Tian S, Luo Y, Zhao Z, Deng N and Ren G 2020 *J. Mol. Model.* 26 55
[54] Herold E, Strauch M, Michalik D, Appelhagen A and Ludwig R 2014 *ChemPhysChem* 15 3040